\documentclass[reprint,superscriptaddress,amsmath,amssymb,aps,pra]{revtex4-1}

\usepackage{graphicx}
\usepackage{dcolumn}
\usepackage{bm}
\usepackage{mathtools,dsfont}
\usepackage{braket}
\usepackage[utf8]{inputenc}
\usepackage{float,comment}
\usepackage{hyperref,xcolor}
\hypersetup{colorlinks=true,linkcolor=blue,citecolor=blue,urlcolor=blue}

\begin{document}

\title{Trapping potentials and quantum gates for microwave-dressed Rydberg atoms\\ on an atom chip}

\author{Iason Tsiamis}
\email{iason.tsiamis@fu-berlin.de}
\affiliation{Dahlem Center for Complex Quantum Systems and Fachbereich Physik, Freie Universität Berlin, 14195 Berlin, Germany}
\affiliation{Institute of Electronic Structure and Laser and Center for Quantum Science and Technologies, FORTH, 70013 Heraklion, Crete, Greece}

\author{Georgios Doultsinos}
\affiliation{Institute of Electronic Structure and Laser and Center for Quantum Science and Technologies, FORTH, 70013 Heraklion, Crete, Greece}
\affiliation{Department of Physics, University of Crete, Heraklion, Greece}

\author{Andreas F. Tzortzakakis}
\affiliation{Embedded Intelligence, German Research Center for Artificial Intelligence, 67663 Kaiserslautern, Germany}

\author{Manuel Kaiser}
\affiliation{Center for Quantum Science, Physikalisches Institut, Eberhard Karls Universität Tübingen, Auf der Morgenstelle 14, 72076 Tübingen, Germany}

\author{Dominik Jakab}
\affiliation{Center for Quantum Science, Physikalisches Institut, Eberhard Karls Universität Tübingen, Auf der Morgenstelle 14, 72076 Tübingen, Germany}

\author{Andreas Günther}
\affiliation{Center for Quantum Science, Physikalisches Institut, Eberhard Karls Universität Tübingen, Auf der Morgenstelle 14, 72076 Tübingen, Germany}

\author{József Fortágh}
\affiliation{Center for Quantum Science, Physikalisches Institut, Eberhard Karls Universität Tübingen, Auf der Morgenstelle 14, 72076 Tübingen, Germany}

\author{David Petrosyan}
\affiliation{Institute of Electronic Structure and Laser and Center for Quantum Science and Technologies, FORTH, 70013 Heraklion, Crete, Greece}
\affiliation{Center for Quantum Science, Physikalisches Institut, Eberhard Karls Universität Tübingen, Auf der Morgenstelle 14, 72076 Tübingen, Germany}


\begin{abstract}
Rydberg atoms in dc electric fields acquire static dipole moments.  
When the atoms are close to a surface producing an inhomogeneous electric field,  such as by the adsorbates on an atom chip, depending on the sign of the dipole moment of the Rydberg-Stark eigenstate, the atoms may experience a force towards or away from the surface. 
We show that by applying a bias electric field and coupling a desired Rydberg state by a microwave field of proper frequency to another Rydberg state with opposite sign of the dipole moment, we can create a trapping potential for the atom at a prescribed distance from the surface. 
Perfectly overlapping trapping potentials for several Rydberg states can also be created by multi-component microwave fields. 
A pair of such trapped Rydberg states of an atom can represent a qubit. 
Finally, we discuss an optimal realization of the $\textsc{swap}$ gate between pairs of such atomic Rydberg qubits separated by a large distance but interacting with a common mode of a planar microwave resonator at finite temperature.
\end{abstract}

\maketitle

\section{Introduction}
\label{Sec:Intro}

Integrated superconducting atom chips provide a versatile platform to realize hybrid quantum systems enabling coherent coupling between cold atoms and superconducting circuits \cite{Xiang2013,Kurizki2015,Lauk2020}.
Achieving strong interactions between atoms and on-chip planar microwave resonators requires positioning the atoms near the chip surface and exciting them to Rydberg states having strong electric dipole transitions of appropriate frequencies \cite{Sorensen2004,Petrosyan2009,Hogan2012,Kaiser2022}. 
However, inhomogeneous static electric fields from adsorbates on the chip surface induce spatially varying energy shifts of Rydberg states \cite{Tauschinsky2010,Hattermann2012}, complicating their coherent manipulation for quantum information applications. 
Moreover, the spatial inhomogeneity of the field can lead to sizable mechanical force acting on Rydberg atoms, resulting in decoherence or even their loss.

Various strategies have been explored to suppress adsorbate fields: 
Using specially oriented quartz crystals as substrate materials can reduce adsorbate binding \cite{Sedlacek2016}, 
while covering the chip surface with a thin metallic (rubidium) layer neutralizes stray electric fields, enabling coherent Rydberg state manipulation \cite{Hermann-Avigliano2014}. 
Microwave dressing further reduces differential Stark shifts between Rydberg state pairs, enhancing their coherence \cite{Jones2013}. 
Despite these advances, mitigating the effects of inhomogeneous adsorbate fields remains an active area of research.

As an alternative to suppressing adsorbate fields, here we exploit the typical conditions on atom chips to trap and coherently manipulate Rydberg atoms. 
Inspired by the methods to trap paramagnetic atoms in radio-frequency adiabatic potentials \cite{Perrin2017}, we demonstrate that one can create a trapping potential for a Rydberg atom in an inhomogeneous electric field using a microwave field near-resonant with an electric dipole transition between a pair of Rydberg-Stark eigenstates with opposite sign of the static dipole moment. 

Our method enables trapping Rydberg atoms in principle in any spatially varying electric fields, but we focus here on trapping Rydberg atoms near the surface of an integrated superconducting atom chip for quantum information applications.   
By tuning the microwave frequency and using an additional homogeneous bias field, the trapping potential can be placed at a precise distance from the chip surface. 
Moreover, well-aligned trapping potentials for several Rydberg states can be created by multi-component microwave fields. 
Hence, a pair of trapped Rydberg states of an atom can represent a qubit driven by a resonant microwave field or interacting with a particular microwave mode of the on-chip planar waveguide resonator. 
Then, pairs of such qubits near the chip surface separated from each other by large distances but coupled simultaneously to the same cavity mode can interact via exchange of virtual photons \cite{Petrosyan2008,Sarkany2015,Sarkany2018}. This can realize the \textsc{swap} operation \cite{Sarkany2018} or the entangling $\sqrt{\textsc{swap}}$ gate between the Rydberg qubits mediated by the cavity even at finite temperature and thus containing thermal photons.

\section{Microwave-dressed potentials for Rydberg atoms}
\label{Sec:trap}

\begin{figure}[t]
    \includegraphics[clip,width=1.0\columnwidth]{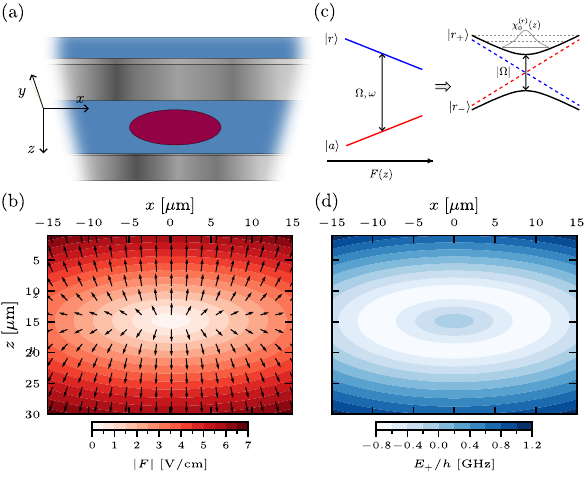}
    \caption{(a)~Illustration of a portion of the integrated atom chip with a patch of adsorbates (red disk) on the surface of dielectric gap (blue) between the superconductors (silver) confining the resonator field.
      The adsorbates, modelled as a uniformly charged disk (centered at $x,y,z=0$) of radius $83\:\mu$m and charge density $6.45 \times 10^{-20}\:\mathrm{C} / \mu \mathrm{m}^2$, produce a spatially inhomogeneous electric field $F_{\mathrm{ad}}$ which is partially compensated by a homogeneous bias field $F_{\mathrm{b}}$, leading to the total field $F=F_{\mathrm{ad}} + F_{\mathrm{b}}$. 
    (b)~Absolute value of total electric field $|F|$ near the chip surface vs coordinates $x,z$ for $y=0$. The bias field $F_{\mathrm{b}}=-30\:$V/cm fully compensates the adsorbate field in the vicinity of $x,y=0, z=15\:\mu$m. 
      Arrows indicate the mechanical force $\bm{\nabla} (d_r|F|)$ acting on an atom in the Rydberg-Stark state $\ket{r}$ with static dipole moment $d_r>0$.
    (c)~Schematic illustration of the Rydberg atom trapping scheme in one dimension: A pair of atomic Rydberg states $\ket{r}$ and $\ket{a}$ with static dipole moments $d_a\simeq -d_r$ are energy shifted by an electric field $F(z)$ in the opposite directions (left). 
    The atom is irradiated by a microwave field of frequency $\omega$ detuned by $\Delta = \omega - \omega_{ra}$ from the unperturbed atomic resonance $\omega_{ra}$. 
    In a spatially varying field $F(z)$, in the frame rotating with frequency $\omega$, the Rydberg levels cross at $z_0$ such that $\hbar \Delta + (d_r-d_a)|F(z_0)| =0$. 
    The microwave field coupling the levels $\ket{r}$ and $\ket{a}$ with the Rabi frequency $\Omega$ lifts this degeneracy and results in avoided crossing of the two eigenstates $\ket{r_{\pm}}$ with energies $E_{\pm}$ split by $|\Omega|$ (right). 
    The upper potential curve $E_+$ forms a trapping potential with the vibrational ground-state wavefunction $\chi_0^{(r)} (z)$ centered at $z_\mathrm{min} \simeq z_0$.
    (d)~Spatial dependence of the trapping potential $E_+$ near the chip surface vs coordinates $x,z$ for $y=0$, for the static field as in panel (b) and atomic parameters as in Fig.~\ref{fig:Fnuz0}. Gravity is neglected.
    }
    \label{fig:comb}
\end{figure}

Our system is illustrated in Fig.~\ref{fig:comb}.
In Fig.~\ref{fig:comb}(a) we show a portion of an atom chip \cite{Kaiser2022,Bothner2017,Hattermann2017} with a patch of adsorbates on the dielectric surface producing an inhomogeneous electric field $F_{\mathrm{ad}}(x,y,z)$. 
We model the adsorbates as a disk of charges \cite{Bochko2020} choosing the disk radius and density of charges such that the resulting field is close to that measured in our experiment \cite{Kaiser2022}.
In particular, with $x=y=0$ at the center of the disk, the adsorbates on the chip surface produce an inhomogeneous electric field $F_{\mathrm{ad}}(z) \approx F_0 e^{-z/\zeta}$ that decays approximately exponentially with distance $z$ ($\lesssim \zeta$) from the surface.
Using an electrode parallel to the chip surface \cite{Kaiser2022}, we can apply a homogeneous bias field $F_{\mathrm{b}}$ along $z$ to partially compensate the surface field. The total field is then $F = F_{\mathrm{ad}} + F_{\mathrm{b}}$ and its absolute value is shown in Fig.~\ref{fig:comb}(b).
If we place in this field an atom in the Rydberg-Stark eigenstate $\ket{r}$ with a static dipole moment $d_r$, it will experience a spatially-dependent level shift $\delta E_r = -d_r|F|$ and mechanical force $-\bm{\nabla} E_r$ proportional to the gradient of the potential $E_r$.

To neutralize this force in the vicinity of a particular field value $F$, we select another Rydberg-Stark eigenstate $\ket{a}$ with the static dipole moment $d_a \simeq -d_r$ and apply a microwave field of frequency $\omega$ resonant with the frequency of $\ket{a}\leftrightarrow \ket{r}$ transition between the Stark-shifted levels: $\omega = \omega_{ra} + (d_r - d_a)|F|/\hbar$ [see Fig.~\ref{fig:comb}(c)].
As discussed in more detail below, this results in a trapping potential $E_+$ for the Rydberg atom in the vicinity of the field value $|F| = \hbar \Delta/(d_r - d_a)$ [see Fig.~\ref{fig:comb}(d)].
The atom is trapped only in the direction of the field gradient; in the transverse directions it is free, but there is no force acting on the atom in the plane transverse to the field gradient, and we neglect gravity.

In what follows, we present a more quantitative derivation of the trapping potential for a Rydberg atom in a spatially inhomogeneous electric field
\begin{equation}
    F(z)=F_0 e^{-z/\zeta} + F_{\mathrm{b}} ,
    \label{eq:field}
\end{equation}
as illustrated in Fig.~\ref{fig:Fnuz0}(a).
Close to the chip surface, $z\ll \zeta$, the total field is approximately linear in $z$, $F(z) \approx  F_0 + F_{\mathrm{b}} - F_0 z/\zeta$. 
We thus consider a one-dimensional model that, strictly speaking, is valid only for $x=y=0$.
We note, however, that in any sufficiently small region of space near the chip surface, we can (re)define the $z$ axis along the direction of the local field gradient and assume linearly varying field along $z$ in this region, while in the transverse $x,y$ directions the field is constant and therefore does not exert force on the atoms.

\subsection{Trapping potential for a Rydberg state atom}

\begin{figure}[t]
    \centering
    \includegraphics[clip,width=\columnwidth]{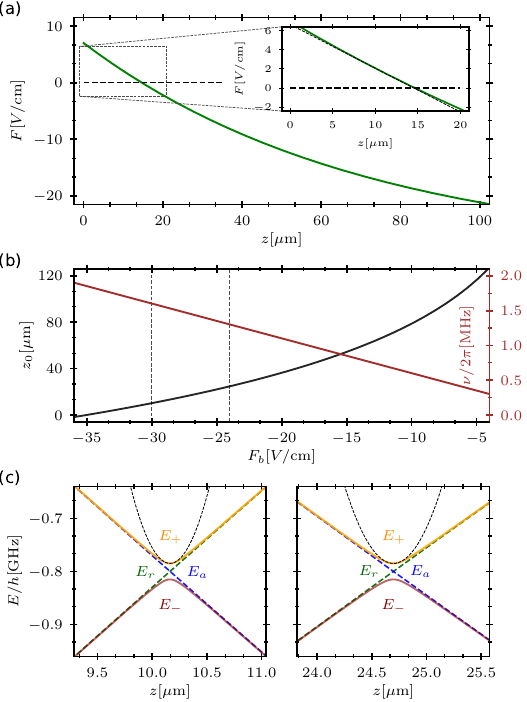}
    \caption{(a) Total electric field $F(z)$ of Eq.~(\ref{eq:field}) vs distance $z$ from the chip.
      The parameters are $F_0=37\:$V/cm, $\zeta=70\:\mu$m \cite{Kaiser2022} and $F_{\mathrm{b}} \simeq -30\:$V/cm.
      Inset: magnified view of the electric field in the linear regime $z\ll \zeta$. 
    (b) Trap position $z_0$ and frequency $\nu$ (for $^{87}$Rb atom) as functions of the applied bias field $F_{\mathrm{b}}$ for fixed $\Delta/2\pi=-1.6\:$GHz, $\Omega/2\pi =30\:$MHz, and $|d_{r,a}|/h \simeq 400\:$MHz/(V/cm). 
    (c) Microwave-dressed energy levels $E_{\pm}$ vs distance $z$ exhibiting avoided crossing at $z_0\simeq 10 \mu$m and $z_0 \simeq 25 \mu$m for $F_{\mathrm{b}}= -30\:$V/cm and $F_{\mathrm{b}}= -24\:$V/cm [vertical dashed lines in panel (b)], respectively. 
    The upper eigenenergy $E_+$ forms a trapping potential approximated by a parabola. }
    \label{fig:Fnuz0}
\end{figure}

Consider a Rydberg atom near the surface of the chip. 
The Rydberg-Stark eigenstates of the atom in a static electric field possess static dipole moments.
We select a Rydberg state $\ket{r}$ and an auxiliary state $\ket{a}$ having dipole moments $d_{r,a}$ that are close in magnitude but have opposite signs, $d_a\simeq -d_r$.
In the static electric field $F$, these two energy levels shift downwards or upwards, $E_{r,a} = \hbar \omega_{r,a} -d_{r,a} |F|$, depending on the sign of their dipole moment, as illustrated in Fig.~\ref{fig:comb}(c). 
We couple level $\ket{r}$ to $\ket{a}$ with a microwave field of frequency $\omega$ near-resonant with the (Stark-shifted) transition frequency $\omega_{ra} - (d_r-d_a)|F|/\hbar$. 
The total Hamiltonian for the atom can be cast as
\begin{equation}
\mathcal{H} = \mathcal{H}_0 + \mathcal{H}_F + \mathcal{H}_{\mathrm{MW}},
\end{equation}
where $\mathcal{H}_0 = \hbar \omega_r \hat{\sigma}_{rr} + \hbar \omega_a \hat{\sigma}_{aa}$ describes the unperturbed Rydberg energy levels,
$\mathcal{H}_F = -d_r |F(z)| \hat{\sigma}_{rr} -d_a |F(z)| \hat{\sigma}_{aa}$ describes the level shifts in the electric field, and 
$\mathcal{H}_{\mathrm{MW}} = -\hbar \frac{1}{2}\Omega e^{-i \omega t} \hat{\sigma}_{ra} + \mathrm{H.c.}$ is the interaction with the microwave field with Rabi frequency $\Omega$ under the rotating-wave approximation, and $\hat{\sigma}_{\mu\nu} \equiv \ket{\mu}\bra{\nu}$ are the atomic projection ($\mu=\nu$) or transition ($\mu\neq \nu$) operators. 
Using the unitary transformation $\mathcal{U} = \exp [-i \omega_r \hat{\sigma}_{rr} t - i (\omega_r -\omega) \hat{\sigma}_{aa} t]$, we obtain the interaction Hamiltonian  in the rotating frame
\begin{equation}
\tilde{\mathcal{H}} = \mathcal{U}^\dag [\mathcal{H} - i \hbar \partial_t] \mathcal{U} = 
\hbar \begin{pmatrix} - d_r |F(z)|/\hbar & -\frac{1}{2} \Omega \\ 
- \frac{1}{2}\Omega^* & \Delta -d_a |F(z)|/\hbar \end{pmatrix}, \label{eq:H2x2}
\end{equation}
where $\Delta=\omega-\omega_{ra}$ is the detuning of the microwave field with respect to the unperturbed atomic resonance $\omega_{ra}$, and for simplicity, we assume linear dc Stark effect with $d_{r,a}$ independent of $F$  (or slowly varying with $F$; see below). 
In the limit of $\Omega\to 0$, the energies $E_{r,a}$ of states $\ket{r},\ket{a}$ cross (in the rotating frame) when 
$\bar{\Delta}(z) \equiv \Delta + (d_r-d_a) |F(z)|/\hbar = 0$, corresponding to distance  
\begin{eqnarray}
  z_0 & = & - \zeta\,\ln \left(\frac{\hbar}{d_a-d_r}\,\frac{\Delta}{F_0} - \frac{F_{\mathrm{b}}}{F_0} \right) \\
 & \approx & \zeta \left(1 - \frac{\hbar}{d_a-d_r}\,\frac{\Delta}{F_0} + \frac{F_{\mathrm{b}}}{F_0}\right) , \nonumber
\end{eqnarray}
where we assumed $F(z) >0$ \cite{commentz0}.
Note that $z_0$ can be tuned by changing the microwave detuning $\Delta$ or varying the bias field $F_{\mathrm{b}}$, as illustrated in Fig.~\ref{fig:Fnuz0}(b). 
Once we switch on the microwave field ($\Omega \neq 0$) that hybridizes the state $\ket{r}$ with $\ket{a}$, the crossing turns into avoided crossing. 
The eigenstates $\ket{r_{\pm}}$ and the corresponding eigenenergies $E_{\pm}$ of Hamiltonian (\ref{eq:H2x2}) are
\begin{subequations}
\begin{eqnarray}
& \ket{r_{\pm}} = \frac{1}{\sqrt{N_\pm}} \Big\{ 
\Big[ \bar{\Delta}(z) \mp \sqrt{ \bar{\Delta}^2(z) + |\Omega|^2} \Big] \ket{r} + \Omega \ket{a} \Big\} , \qquad
\\
& E_\pm (z)/\hbar = \frac{1}{2} \Big[ \bar{\Delta}(z)
\pm \sqrt{ \bar{\Delta}^2(z) + |\Omega|^2} \Big] - d_r |F(z)|/\hbar ,  \qquad \label{eq:pot}
\end{eqnarray}
\end{subequations}
where $N_\pm$ are the normalization factors.
The dressed energy eigenvalues are shown in Fig.~\ref{fig:Fnuz0}(c). In the vicinity of the avoided crossing, $z \simeq z_0$, they are split by $|\Omega|$ and the upper potential curve $E_+(z)$ has a well that can trap the atom at a prescribed distance from the chip surface. 

To characterize the microwave-dressed trapping potential, we expand $E_+(z)$ around the local minimum $z_{\min} = z_0$ as 
\begin{equation}
    E_+\simeq E_+^{\min}+\frac{1}{2}k(z-z_{\min})^2,
    \label{eq:Epluspot} 
\end{equation}
where $E_+^{\min}/\hbar = \Delta d_r/(d_r-d_a) + \frac{1}{2}|\Omega|$ is the potential minimum and 
\begin{equation}\label{eq:springConst}
k = \frac{[\hbar \Delta + (d_r-d_a) F_{\mathrm{b}}]^2}{2\hbar |\Omega| \zeta^2} 
\end{equation}
is the force constant, which reduces to $k \simeq (d_r-d_a)^2 F_0^2/(2\hbar |\Omega| \zeta^2)$ for $z_0 \ll \zeta$. 
For an atom of mass $m_{\mathrm{at}}$, we thus obtain an approximately harmonic trap with vibrational frequency $\nu = \sqrt{k/m_{\mathrm{at}}}$, while the wavefunction of the ground vibrational state ($n=0$)  
\[
\chi_0^{(r)}(z) = \left( \frac{1}{\pi \sigma^2} \right)^{1/4} e^{-\frac{(z-z_{\min})^2}{2\sigma^2}} 
\]
has the width $\sigma  = \sqrt{\hbar /(m_{\mathrm{at}} \nu)}$.
In Fig.~\ref{fig:Fnuz0}(b), we plot the trap position $z_{\min}$ and vibrational frequency $\nu$ versus the bias field $F_{\mathrm{b}}$ for fixed $\Delta$ and $\Omega$. 
We observe that with decreasing the magnitude of $|F_{\mathrm{b}}|$ the trap position $z_{\min} = z_0$ moves away from the chip surface and the trap becomes wider (decreasing $\nu$).

In the above derivation, we assumed linear dc Stark effect for the atoms in the Rydberg states $\ket{r}$ and $\ket{a}$, but for the Rydberg states with low orbital angular momentum $L \leq 3$ and non-vanishing quantum defects, the dc Stark shift is in general nonlinear, meaning that the dipole moment $d_r(F)$ is a function of the electric field $F$. With the spatial extent of the trapped wavefunction $\sigma$, we can estimate that the variation of the electric field $\Delta F \simeq F_0 \sigma /\zeta$ around $F(z_0)$ in the trap is $\Delta F/F < 0.05$. In this interval of fields, the dipole moment can be assumed constant to a very good approximation.

We note that if $|d_r| \neq |d_a|$  (but $d_r$ and $d_a$ still have opposite signs), the position of potential minimum $z_{\min}$ in Eq.~(\ref{eq:Epluspot}) will be slightly shifted from the position of level crossing $z_0$, and the expressions for the trap frequency and width will deviate from the above expressions, as discussed in Appendix~\ref{app:Trap}.     

In Fig.~\ref{fig:Fnuz0}, we illustrate the foregoing discussion using the experimental values for the inhomogeneous adsorbate field \cite{Kaiser2022} and the compensating bias field and employing realistic atomic parameters for an alkaline Rydberg atom. 
In Appendix \ref{app:RydStates}, we show two examples of Stark maps of Rydberg state manifolds of Rb atom suitable for the proposed trapping scheme.

\subsection{Trapping potentials for two Rydberg states}

We can simultaneously create colocalized trapping potentials for other Rydberg states of the atom using additional frequency components of the microwave field to couple the desired Rydberg states to appropriate auxiliary states. 
Consider another Rydberg state $\ket{s}$ having static dipole moment $d_s$. We select an auxiliary state $\ket{b}$ with dipole moment $d_b \simeq -d_s$ and apply an additional microwave field of frequency $\omega'$, near-resonant with the (Stark-shifted) $\ket{b}\leftrightarrow\ket{s}$ transition. The corresponding transition frequency at position $z_0$ is $\omega_{sb} - (d_s-d_b)|F(z_0)|/\hbar$. By setting the detuning 
\begin{equation}
\Delta' \equiv \omega' - \omega_{sb} = -(d_s-d_b) |F(z_0)|/\hbar = \Delta \frac{(d_s-d_b)}{(d_r - d_a)} , 
\label{eq:Deltaprime}
\end{equation}
we ensure that levels $\ket{s}$ and $\ket{b}$ cross again at $z_0$.
Simultaneously, by choosing the Rabi frequency $\Omega'$ of the second microwave field driving the transition $\ket{s} \leftrightarrow \ket{b}$ as 
\begin{equation}
|\Omega'| = |\Omega| \frac{(d_s-d_b)^2}{(d_r-d_a)^2} ,
\end{equation}
we ensure that the trapping potential $E_{+}'$ for $\ket{s_{+}}$
centered at $z_{\min}'=z_0$ has the same vibrational frequency $\nu' = \nu$. Hence, both trapping potentials for $\ket{s_{+}}$ and $\ket{r_{+}}$ have well-overlapping ground-state wavefunctions $\chi_0^{(s)} (z) \simeq \chi_0^{(r)} (z)$.

We note again that if $|d_s| \neq |d_b|$  (but $d_s$ and $d_b$ have opposite signs), $z_{\min}'$ will not coincide with $z_0$ and $z_{\min}$ of the potential $E_{+}$. 
However, this mismatch can be corrected by slightly shifting the detuning $\Delta'$ in Eq.~(\ref{eq:Deltaprime}) and thereby the level crossing point $z_0'$, which will permit to perfectly align the potential minima $z_{\min}' = z_{\min}$ (see Appendix~\ref{app:Trap}).     
Examples of suitable Rydberg states $\ket{s}$ and $\ket{b}$ (as well as $\ket{r}$ and $\ket{a}$) of Rb are shown in Appendix \ref{app:RydStates}. 

\subsection{Rydberg atom qubit}

We can thus use a pair of trapped Rydberg states of the atom to represent a qubit with the basis states $\ket{0} \equiv \ket{s_+} \otimes \chi_0^{(s)}$ and $\ket{1} \equiv \ket{r_+} \otimes \chi_0^{(r)}$ and the corresponding Bohr frequencies $\omega_0 = \omega_{s} - d_s |F(z_0)|/\hbar + \frac{1}{2} \Omega' +\frac{1}{2} \nu'$ and $\omega_1 = \omega_{r} - d_r |F(z_0)|/\hbar + \frac{1}{2} \Omega +\frac{1}{2} \nu$. 
The Rydberg qubit transition $\ket{0} \leftrightarrow \ket{1}$ can be driven by a near-resonant microwave field $\varepsilon_q$ of frequency $\omega_q \simeq \omega_{10} \equiv \omega_1 - \omega_0$.  
Given the dipole matrix element $\wp_{sr}$ for the Rydberg transition $\ket{s} \to \ket{r}$, the transition $\ket{0} \leftrightarrow \ket{1}$ has the dipole matrix element $\wp_{01} = \frac{1}{2} f_{sr} \wp_{sr}$, where the factor $1/2$ comes from $\bra{s_+} \hat{\sigma}_{sr} \ket{r_+} \simeq 1/2$ and the Franck-Condon factor $f_{sr}$ is proportional to the overlap of the spatial wavefunctions, 
\[
f_{sr}=\int \chi_0^{(s)*}(z) \chi_0^{(r)}(z) dz = 
\left( \frac{2 \sigma \sigma'}{\sigma^2 + \sigma^{'2}}  \right)^{1/2} e^{-\frac{(z_{\min}-z_{\min}')^2}{2(\sigma^2 + \sigma^{'2})}} , 
\]
which reduces to $f_{sr}\simeq 1$ for $z_{\min} \simeq z_{\min}'$ and $\nu \simeq \nu'$. 
The Rabi frequency of the microwave field driving the qubit transition is $\Omega_q = \wp_{01} \varepsilon_q/2\hbar$, and a resonant, $\omega_q = \omega_{10}$, microwave pulse of area $\theta = \int \Omega_q dt =\pi$ will flip the qubit state $\ket{0} \leftrightarrow \ket{1}$, while a $\pi/2$ pulse applied to a qubit in, e.g., state $\ket{0}$ will create a superposition state $\ket{0} \to (\ket{0}+ie^{i\phi} \ket{1})/\sqrt{2}$, where $\phi$ is the phase of the microwave field. 
Hence, arbitrary qubit rotations $R_{x,y}(\theta)$ about the $\hat{x}$ and $\hat{y}$ axes of the Bloch sphere can be performed by a resonant microwave field of proper area $\theta$ and phase $\phi=-\pi/2,0$.  
Rotations $R_{z}(\theta)$ about the $\hat{z}$ axis can be induced by, e.g., a non-resonant microwave field, $|\Delta_q| \gg \Omega_q$ ($\Delta_q = \omega_q - \omega_{10}$), inducing level shifts $\delta\omega_{0,1} = \mp |\Omega_q|^2/2\Delta_q$ of $\ket{0}$ and $\ket{1}$ in the opposite directions, and thereby their phase shift $\theta = \delta\omega_{0,1} t$ during time $t$.      

Note that since the Rydberg states of atoms decay, the Rydberg qubits are, in general, not suitable for long-term storage of quantum information. 
We therefore assume that qubit states are stored as superpositions of the long-lived hyperfine sublevels $\ket{g_0}$ and $\ket{g_1}$ of the electronic ground state of an atom trapped in optical tweezers at position $z=z_0$ near the chip surface \cite{footnote}.
When we need to perform quantum gates with Rydberg qubits, we prepare the trapping states $\ket{r_+}$ and $\ket{s_+}$ by turning on the microwave fields $\Omega^{(\prime)}$ in the presence of appropriate bias field, and then transfer the atom from the storage states $\ket{g_0}$ and $\ket{g_1}$ to the Rydberg qubits states   
$\ket{0}$ and $\ket{1}$ using resonant lasers. 
A laser with carrier frequency $\omega_{L_0} = \omega_0-\omega_{g_0}$ and pulse area $\pi$ will then transfer $\ket{g_0} \to \ket{0}$, and 
a laser with carrier frequency $\omega_{L_1} = \omega_1-\omega_{g_1}$ and pulse area $\pi$ will transfer $\ket{g_1} \to \ket{1}$. 
With realistic (two- or three-photon) Rabi frequencies $\Omega_L/2\pi \gtrsim 1\;$MHz, the duration of the laser $\pi$ pulses is $\tau \lesssim 1\:\mu$s.    
Once the quantum gates are performed (see below), the Rydberg qubits are transferred to the storage states by another fast $\pi$ pulse.

\section{Cavity-mediated SWAP gate between Rydberg-atom qubits}

To implement quantum gates between the trapped Rydberg qubits, we can couple them to a common mode of an on-chip microwave resonator that will mediate their interaction \cite{Sarkany2018}. 
We thus consider a pair of atoms $j=1,2$ placed at different but equivalent positions within the electric field mode of the microwave cavity. 
Meanwhile, the atoms are largely separated from each other and direct interaction between their Rydberg states is negligibly small. 
The Hamiltonian for the system is 
\begin{equation}
\mathcal{H} = \mathcal{H}_c + \sum_j (\mathcal{H}_{q}^{(j)} +  \mathcal{V}_{qc}^{(j)}) ,  
\end{equation}
where $\mathcal{H}_c = \hbar \omega_c \hat{c}^\dag\hat{c}$ describes the cavity mode with frequency $\omega_c$ and the photon creation and annihilation operators $\hat{c}^\dag,\hat{c}$, $\mathcal{H}_q^{(j)} = \hbar \omega_0^{(j)} \hat{\sigma}_{00}^{(j)} + \hbar \omega_{1}^{(j)} \hat{\sigma}_{11}^{(j)}$ describes the qubit levels of atom $j$, and $\mathcal{V}_{qc}^{(j)} = g_i \hat{\sigma}_{10}^{(j)} \hat{c} + \mathrm{H.c.}$ is the qubit-cavity interaction in the rotating wave approximation. The coupling strength $g_j = (\wp_{01}/\hbar) \varepsilon_c u(\mathbf{r}_j)$ is proportional to the $\ket{0} \leftrightarrow \ket{1}$ transition dipole moment $\wp_{01}$, the field per photon $\varepsilon_c = \sqrt{\hbar \omega_c/\epsilon_0 V_c}$ in the cavity with the effective volume $V_c$, and the cavity mode function $u(\mathbf{r}_j)$ at the atomic position $\mathbf{r}_j$.
For a coplanar waveguide resonator with the stripline length $L$ and the grounded electrodes at distance $D$ (dielectric gap), the effective cavity volume is $V_c =2\pi D^2 L$ and the mode function near the standing wave field antinode falls off with distance from the surface as $u(\mathbf{r}) \simeq e^{-z/D}$ \cite{Verdu2009}. 

Using the unitary transformation 
$\mathcal{U} = \exp[-i \omega_c \hat{c}^\dag\hat{c} t] \otimes 
\exp[-i \sum_j (\omega_0^{(j)} + \delta_c^{(j)}/2) \hat{\sigma}_{00}^{(j)} t ] \otimes 
\exp[-i \sum_j (\omega_1^{(j)} - \delta_c^{(j)}/2) \hat{\sigma}_{11}^{(j)} t  ] $, we obtain the interaction Hamiltonian $\tilde{\mathcal{V}}_{qc} = \mathcal{U}^\dag [\mathcal{H} - i \hbar \partial_t] \mathcal{U}$ in the frame rotating with the frequency of the cavity mode:
\begin{equation}
    \tilde{\mathcal{V}}_{qc} = \sum_{j=1,2} \left[ \tfrac{1}{2} \delta_c^{(j)} \left( \hat{\sigma}_{00}^{(j)} - \hat{\sigma}_{11}^{(j)} \right) + g_j \left( \hat{c}^\dag \hat{\sigma}_{01}^{(j)} + \hat{\sigma}_{10}^{(j)} \hat{c} \right) \right] , \label{eq:AtCavInt}
\end{equation}
where $\delta_c^{(j)} = \omega_c - \omega_{10}^{(j)}$ is the detuning of the cavity mode from the qubit transition.

Several relaxation processes affect the system. 
We assume that the Rydberg qubit states $\ket{0(1)}_j$ of each atom irreversibly decay to state(s) $\ket{l}_j$ outside the computational space with rate $\gamma$, which is described by the Lindblad operators $\hat{L}_{0(1)}^{(j)} = \sqrt{\gamma} \hat{\sigma}_{l0(1)}^{(j)}$.  
Then, the cavity field relaxes with rate $\kappa$ toward thermal equilibrium with the mean number  of thermal photons $\bar{n}_{\mathrm{th}} = (e^{\hbar \omega_c/k_B T} - 1)^{-1}$ at temperature $T$ as described by Lindblad operators $\hat{L}_{-}^{(c)} =\sqrt{\kappa (\bar{n}_{\mathrm{th}} +1)} \hat{c}$ and $\hat{L}_{+}^{(c)} = \sqrt{\kappa \bar{n}_{\mathrm{th}}} \hat{c}^\dag$ \cite{PLDP2007}.

We simulate the dissipative dynamics of the system using the quantum Monte Carlo stochastic wavefunction approach \cite{Molmer1993,Plenio1998,PLDP2007}. 
We thus propagate the wavefunction $\ket{\Psi}$ of the compound system, consisting of the two atoms $j=1,2$ with the internal states $\ket{0,1,l}_{j}$ and the cavity field containing $n \in [0,n_{\max} ]$ photons (with a sufficiently large cut-off $n_{\max} = 10 \bar{n}_{\mathrm{th}}$) with the effective non-Hermitian Hamiltonian 
\begin{equation}
    \mathcal{H}_{\mathrm{eff}} = \tilde{\mathcal{V}}_{qc} 
    - \frac{i \hbar}{2} \Big[ \sum_{j=1,2} \sum_{\mu=0,1} \hat{L}_{\mu}^{(j)\dag} \hat{L}_{\mu}^{(j)} + \sum_{\nu=-,+} \hat{L}_{\nu}^{(c)\dag} \hat{L}_{\nu}^{(c)} \Big]  \label{eq:HameffnH}
\end{equation}
that does not preserve the norm $\braket{\Psi|\Psi}$ of the wavefunction.   
The evolution is accompanied by quantum jumps $\ket{\Psi} \to \hat{L}_{\alpha}^{(\beta)} \ket{\Psi}/ \sqrt{\bra{\Psi} \hat{L}_{\alpha}^{(\beta)\dag} \hat{L}_{\alpha}^{(\beta)} \ket{\Psi}}$ along the different relaxation channels with probabilities proportional to $\bra{\Psi} \hat{L}_{\alpha}^{(\beta)\dag} \hat{L}_{\alpha}^{(\beta)} \ket{\Psi}$. 
Hence, decay of the atomic Rydberg states $\ket{0(1)}_j \to \ket{l}_j$ results in qubit loss, while quantum jumps between the different photon number states $\ket{n} \to \ket{n \pm 1}$ lead to decoherence of the two-qubit dynamics, as discussed below. 

\subsection{Optimal detuning}

\begin{figure}[t]
\includegraphics[clip,width=\columnwidth]{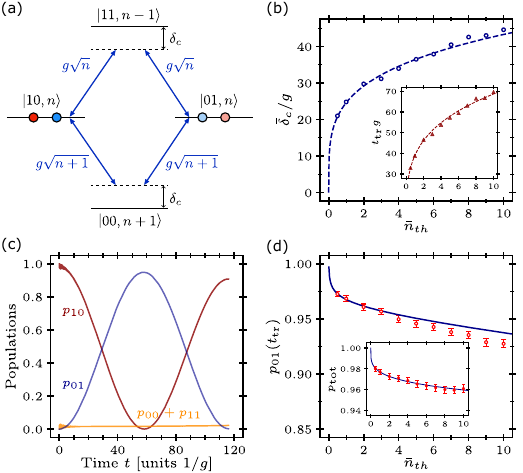}
\caption{(a) Transition paths between the states $\ket{10,n}$ and $\ket{01,n}$ involving the intermediate states $\ket{00,n+1}$ and $\ket{11,n-1}$ containing one more or one less cavity photon.
(b)~Optimal detuning $\bar{\delta}_c$ vs the mean thermal photon number $\bar{n}_{\mathrm{th}}$. 
Inset shows the corresponding \textsc{swap} time $t_\mathrm{tr} \approx \frac{\pi}{2} \frac{\bar{\delta}_c}{g^2}$.  
(c)~Illustration of oscillation dynamics between the states $\ket{01}$ and $\ket{10}$ for one full Rabi cycle $0<t<2t_{\mathrm{tr}}$ for $\bar{n}_{\mathrm{th}}=5$ as obtained from Monte Carlo simulations.
(d)~Population $p_{01}$ of the target state $\ket{01}$ at time $t_\mathrm{tr}$ vs $\bar{n}_{\mathrm{th}}$ as obtained analytically for $\kappa=0$ and $\gamma = 3 \times 10^{-4}g$, and numerically for $\kappa=10^{-3}g$ via the Monte Carlo simulations involving 5000 independent trajectories for each data point (error bars indicate one standard deviation). 
Inset shows the sum of populations of the atomic states, $p_\mathrm{tot} = p_{00}+p_{01}+p_{10}+p_{11}$.}
\label{fig:AtCavInt}
\end{figure}

We assume that the two atoms are trapped in equivalent positions in the cavity: $g_{1,2} = g$ and $\delta_c^{(1,2)} = \delta_c$. 
Our aim is to implement a two-qubit i\textsc{swap} gate that swaps the states $\ket{10}$ and $\ket{01}$ ($\ket{10} \leftrightarrow i \ket{01}$) and leaves the states $\ket{00}$ and $\ket{11}$ unchanged. 
The excitation exchange between the qubits is mediated by the cavity mode. 
Consider the states $\ket{10,n}$ and $\ket{01,n}$ coupled by the atom-cavity interaction Hamiltonian (\ref{eq:AtCavInt}) to states $\ket{00,n+1}$ and $\ket{11,n-1}$ [see Fig.~\ref{fig:AtCavInt}(a)].
In order to suppress the evolution (population or depopulation) of states $\ket{00}$ and $\ket{11}$ and to minimize the effects of relaxation and thermalization of the cavity mode during the transfer, the atoms should exchange a virtual cavity photon.
We therefore should choose sufficiently large detuning $|\delta_c| \gg g\sqrt{n_{\max}}$, where $n_{\max}$ is the maximum number of photons that can be in the cavity with appreciable probability; for a thermal cavity with the photon number probability distribution $P_{\bar{n}_\mathrm{th}} (n) = \bar{n}_{\mathrm{th}}^n/ (1+\bar{n}_{\mathrm{th}})^{n+1}$, we can assume $n_{\max} \lesssim 10 \, \bar{n}_{\mathrm{th}}$.
Then, using perturbation theory to adiabatically eliminate the nonresonant states $\ket{00,n+1}$ and $\ket{11,n-1}$, we obtain the second-order transition amplitude between states $\ket{10,n}$ and $\ket{01,n}$ as 
\begin{equation}
    G(n) \approx - \frac{g^2(n+1)}{\delta_c} + \frac{g^2n}{\delta_c} = - \frac{g^2}{\delta_c} ,
\end{equation}
where the first and second terms correspond to the amplitudes of the transition via the lower and upper paths in Fig.~\ref{fig:AtCavInt}(a), 
and we neglected the higher-order dependence $\propto \frac{g^4}{\delta_c^3}n^2$ of these amplitudes on $n$, which partially cancel in their sum in $G(n)$ \cite{Sarkany2018}. 
Hence, the exchange rate $G$ is nearly independent on $n$ and this approximation becomes even better for larger detunings $\delta_c$.
However, larger detuning leads to slower excitation exchange between the two qubits, $\ket{10} \to \cos (Gt) \ket{10} - i \sin(Gt) \ket{01}$ (or $\ket{01} \to \cos (Gt) \ket{01} - i \sin(Gt) \ket{10}$), and the longer excitation transfer or \textsc{swap} time $t_{\mathrm{tr}} = \pi/2G$ will lead to larger decay probability of the atomic qubit states, $1-e^{-2\gamma t_{\mathrm{tr}}} \simeq 2\gamma t_{\mathrm{tr}}$. For smaller photon numbers, we can indeed choose smaller detuning, but with increasing $\bar{n}_\mathrm{th}$, we should also increase $\delta_c$ to avoid populating states $\ket{00}$ and $\ket{11}$. 

We therefore proceed to optimize the detuning $\delta_c$ for different $\bar{n}_\mathrm{th}$ as follows. 
Starting with the initial state $\ket{10,n}$ with any $n$ and some $\delta_c$, and neglecting for the moment the cavity photon relaxations, $\kappa = 0$, we use the effective Hamiltonian (\ref{eq:HameffnH}) to solve analytically for the dynamics of the system, $\partial_t \ket{\Psi_n} = -\frac{i}{\hbar} \mathcal{H}_{\mathrm{eff}} \ket{\Psi_n}$,  
obtaining the state vector of the four-level system
$\ket{\Psi_n(t)} = c_{10,n}(t) \ket{10,n} + c_{00,n+1}(t) \ket{00,n+1} + c_{11,n-1}(t) \ket{11,n-1} + c_{01,n}(t) \ket{01,n}$.
Its norm $\braket{\Psi_n|\Psi_n}$ is not preserved but decays exponentially in time with rate $2\gamma$ since we assumed that each qubit level of each atom decays with the rate $\gamma$, while $\kappa = 0$.  
Next, given the mean number of thermal photons in the cavity $\bar{n}_{\mathrm{th}}$, the population of the target state $\ket{01}$ is given by $p_{01} (t) = \sum_n P_{\bar{n}_{\mathrm{th}}}(n) |c_{01,n}(t)|^2$.
Maximizing $p_{01}(t)$ with respect to $\delta_c$ and $t$, we find the optimal detuning $\bar{\delta}_{c} (\bar{n}_{\mathrm{th}})$ and the corresponding transfer time $t_{\mathrm{tr}} \approx \frac{\pi}{2} \frac{\bar{\delta}_c}{g^2}$, as shown in Fig.~\ref{fig:AtCavInt}(b). 
The obtained detuning $\bar{\delta}_{c}$ is optimal for given $\bar{n}_{\mathrm{th}}$ in the sense that larger detuning would result in slower dynamics and more atomic decay, while smaller detuning $\delta_c$, being comparable to $g\sqrt{n}$, would result in larger population of states $\ket{00}$ and $\ket{11}$. 
We verified these conclusions via exact numerical simulations of the dynamics of the system.

Next, we include also the cavity field relaxations, $\kappa > 0$, and perform Monte Carlo simulations of the dissipative dynamics of the system. 
As an illustration, in Fig.~\ref{fig:AtCavInt}(c) we show the dynamics of atomic populations for a certain $\bar{n}_{\mathrm{th}} > 0$ until time $t=2t_{\mathrm{tr}}$, as obtained from averaging over large number of independent quantum trajectories. 
For each trajectory, we start with the state $\ket{10,n}$ with $n$ chosen at random with the probability determined by the equilibrium thermal distribution $P_{\bar{n}_\mathrm{th}} (n)$ for given $\bar{n}_\mathrm{th}$ at temperature $T$. 
The state vector of the system then evolves under the non-Hermitian Hamiltonian $\mathcal{H}_{\mathrm{eff}}$ of Eq.~(\ref{eq:HameffnH}) accompanied by quantum jumps of the atoms to the passive state $\ket{l}_j$ resulting in the qubit loss and termination of the dynamics, or quantum jumps of the cavity photon number between neighboring $n$ resulting in dephasing and accumulation of population in states other than the target state [see Fig.~\ref{fig:AtCavInt}(d)]. 
On average, during time $t_{\mathrm{tr}}$ we obtain $2\kappa \bar{n}_{\mathrm{th}}(\bar{n}_{\mathrm{th}}+1) t_{\mathrm{tr}}$ quantum jumps of the cavity photon number, and for $\bar{n}_{\mathrm{th}} >1$ most of the atomic population missing from the target state $\ket{01}$ and not decayed to passive states $\ket{l}$ is accumulated in $p_{00}$ corresponding to states $\ket{00,n}$ with $n\simeq \bar{n}_{\mathrm{th}}$ favored by the quantum jumps.  
We thus see that even for a thermal cavity containing on average many photons $\bar{n}_{\mathrm{th}} >0$, by using the corresponding optimal detuning $\bar{\delta}_{c}$ we can still obtain rather large transfer probability $p_{01}(t_{\mathrm{tr}})$.

\subsection{Gate fidelity}

\begin{figure}[t]
\includegraphics[clip,width=\columnwidth]{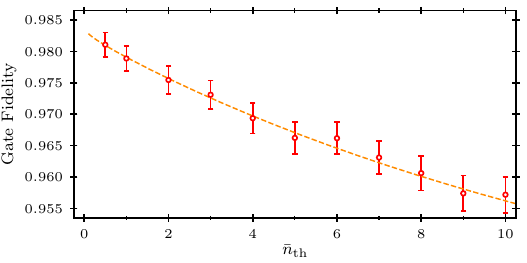}
\caption{Fidelity of preparation of the two-atom entangled state $(\ket{10} - i \ket{01})/\sqrt{2}$ via the $\sqrt{\textsc{swap}}$ gate during time $t_\mathrm{tr}/2$ vs $\bar{n}_{\mathrm{th}}$, as obtained from Monte Carlo simulations averaged over $M=5000$ independent trajectories, with the parameters as in Fig.~\ref{fig:AtCavInt}(c). 
Error bars correspond to one standard deviation and the dashed line is guide for the eye.} 
    \label{fig:Fsqrtswap}
\end{figure}

The excitation transfer, or \textsc{swap}, between two qubits is not an entangling operation. Rather, the $\sqrt{\textsc{swap}}$ is a universal entangling gate that results from the same dynamics as for the transfer but interrupted at time $t_{\sqrt{\textsc{swap}}} = t_{\mathrm{tr}} /2$. 
Starting from state $\ket{10}$, an ideal $\sqrt{\textsc{swap}}$ would result in the preparation of the Bell-like state $\ket{B} = (\ket{10} - i \ket{01})/\sqrt{2}$. 
We thus calculate the fidelity of preparation of such a state as 
\begin{equation}
 \mathcal{F} = \frac{1}{M}\sum_m^M \braket{\Psi^{(m)} (t_\mathrm{tr}/2)| B}  \braket{B|\Psi^{(m)} (t_\mathrm{tr}/2 )} , 
\end{equation}
where $\ket{\Psi^{(m)}}$ is the normalized wave vector for the $m$th quantum trajectory and we sum over many ($M \gg 1$) quantum trajectories.  
The resulting fidelity of the $\sqrt{\textsc{swap}}$ gate is shown in Fig.~\ref{fig:Fsqrtswap} and is well above 0.95 for $\bar{n}_{\mathrm{th}}$ up to 10. 

We finally note that to initiate and terminate the cavity-mediated interactions between the atoms, they should be quickly transferred from the ground storage states to the Rydberg qubit states and back to the storage states by strong resonant laser pulses with Rabi frequencies $G \ll \Omega_L \ll \min |E_+ - E_-| =\Omega$.
Alternatively, a pair of atoms in Rydberg states can be decoupled from each other if their qubit transition frequencies $\omega_{10}^{(j)}$ differ by more than the exchange interaction strength, $|\omega_{10}^{(1)} -  \omega_{10}^{(2)}| \gg G$, which can be achieved by, e.g., inducing ac Stark shift of the Rydberg level(s) of one of the atoms by a focused nonresonant laser field.

\section{Conclusions}

To summarize, we have shown how to form trapping potentials for Rydberg states of atoms in inhomogeneous electric fields using appropriate microwave-field dressing.
Motivated by quantum information applications, we considered trapping Rydberg atoms close to the surface of an integrated superconducting atom chip containing a planar waveguide microwave resonator that can mediate interactions and quantum gates between distant atomic qubits. 
A typical stripline cavity has a length $L \simeq 1\:$cm that determines the maximal distance over which it can mediate the interactions. 
Rydberg atoms placed at $z_0 \lesssim 10\:\mu$m distance from the chip surface near the electric field antinode of the cavity can strongly interact with the cavity field. 
Due to the huge transition dipole moment between the neighboring Rydberg states $\wp_{sr} \sim n_{\mathrm{pqn}}^2 a_0e$, where $n_{\mathrm{pqn}}=50-60$ is the (effective) principal quantum number of the Rydberg states, $a_0$ is the Bohr radius, and $e$ the electron charge, and strong confinement of the field in the small volume $V_c \ll \lambda_c^3$, where $\lambda_c \sim L$ is the wavelength of the near-resonant microwave field mode of the cavity, the atom-field coupling constant (vacuum Rabi frequency) can be very large, $g/2\pi = 5-10\:$MHz, reaching the strong-coupling regime, $g \gg \kappa, \gamma$. 
Then, the cavity can mediate interactions between the atoms exchanging a virtual cavity photon while remaining largely immune to the thermal photons present in the microwave cavity in cryogenic conditions.
While this cavity-mediated interaction scheme is similar to that in our previous work \cite{Sarkany2018} adapted to the trapped Rydberg qubits, instead of actively cooling the cavity using an auxiliary atomic ensemble and wave mixing, we here used a different approach to achieve higher fidelities of quantum gate operations by optimizing the cavity detuning for its given temperature and/or mean thermal photon number.
Our results indicate that the achievable gate fidelities can be increased by using colder resonators with higher quality factors.  
By enabling trapping of Rydberg atoms near an integrated superconducting atom chip involving a microwave resonator, our scheme can thus serve to build a large-scale quantum computer or simulator, where cavity-mediated interactions between any pair of qubits or spins separated by macroscopic distances can be switched at will, thereby complementing the shorter-range dipole-dipole or van der Waals interactions between the Rydberg atoms in free space. 

\begin{acknowledgments}
We thank Jens Eisert for valuable discussions. This work was supported by the EU HORIZON-RIA project EuRyQa (grant No. 101070144), EU QuantERA project MOCA (DFG grant No. 491986552), and DFG FOR 5413/1 project QUSP (grant No. 465199066).
I.T. acknowledges support by DFG through the Emmy Noether program (Grant No. ME 4863/1-1).

\section*{Data availability}The data supporting the results of this work are openly available \cite{tsiamis_data_2025}. 
\end{acknowledgments}

\renewcommand{\appendixname}{APPENDIX}
\appendix

\section{TRAP MINIMA AND WIDTHS}
\label{app:Trap}

In the main text, we assumed the static dipole moments for the Rydberg states $\ket{r}$ and $\ket{a}$ to have the same magnitude but opposite sign, $d_r = -d_a$, leading to the trap position $z_{\min} = z_0$ with the force constant $k = [\hbar \Delta + (d_r-d_a) F_{\mathrm{b}}]^2/(2\hbar |\Omega| \zeta^2)$. 
Here we consider a more general case of $|d_r| \neq |d_a|$, but $\mathrm{sgn}(d_r) = - \mathrm{sgn}(d_a)$, and determine the trap parameters. 

Starting from Eq.~(\ref{eq:pot}), the minimum of the trapping potential $z_{\min}$ is found from $\partial_z E_+(z) = 0$ and its force constant $\tilde{k}$ is obtained by expanding $E_+(z)$ up to second order in $z$ around $z_{\min}$. 
We thus obtain a harmonic potential $E_+\simeq \tilde{E}_+^{\min}+\frac{1}{2} \tilde{k} (z-z_{\min})^2$ with
\begin{eqnarray}
    \tilde{k} &=& k - \frac{d_r + d_a}{2\zeta^2}\left(\frac{\hbar\Delta}{d_r - d_a} + F_{\mathrm{b}}\right) , \\
    z_{\min} &=& z_0 + \frac{d_r + d_a}{2\zeta \tilde{k}}\left(\frac{\hbar\Delta}{d_r - d_a} + F_{\mathrm{b}}\right) , \label{eqapp:z0} \\
    \tilde{E}_+^{\min} &=& E_+^\mathrm{min} -\frac{(d_r +d_a)^2}{8\tilde{k}\zeta^2}\left( \frac{\hbar\Delta}{d_r -d_a} +F_{\mathrm{b}}\right)^2 . \label{eqapp:Epmin}
\end{eqnarray}
For $d_r + d_a = 0$ these equations obviously reduce to those in the main text, while for $d_r \approx-d_a$ we can approximate Eqs.~(\ref{eqapp:z0}) and (\ref{eqapp:Epmin}) to first order in $|d_r +d_a| \ll |d_{r,a}|$ by replacing $\tilde{k}\rightarrow k$.

\section{RYDBERG STATE MANIFOLD}
\label{app:RydStates}

\begin{figure}[t]
    \centering
    \includegraphics[clip,width=\columnwidth]{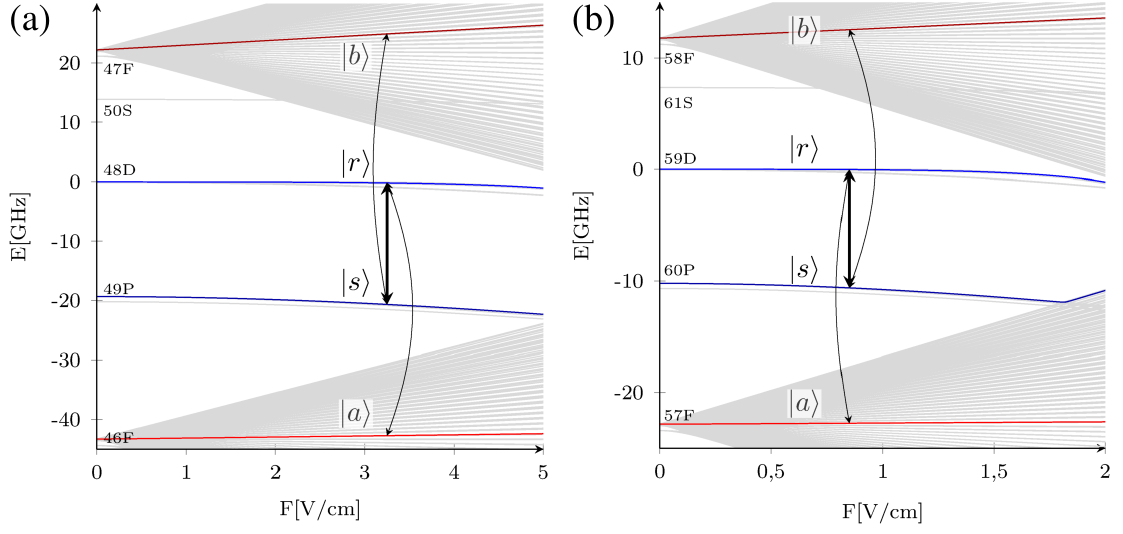}
    \caption{Stark maps of suitable Rydberg state manifolds of an Rb atom in weak electric fields $F$ in the vicinity of (a) $48D$ and (b) $59D$ states. 
    Thin lines with small arrows represent the microwave-field dressing of states $\ket{r}$ with $\ket{a}$ and $\ket{s}$ with $\ket{b}$ for trapping. 
    The thicker line connecting states $\ket{r}$ and $\ket{s}$ represents either a microwave field driving the qubit transition with Rabi frequency $\Omega_q$ or the cavity field coupled to the qubit transition with strength $g$.}
    \label{fig:StarkMaps}
\end{figure}

In Fig.~\ref{fig:StarkMaps}, we show Stark maps of Rydberg state manifolds of an Rb atom that can be used for trapping and qubit manipulations discussed in the main text. 

In the first example [Fig.~\ref{fig:StarkMaps}(a)], we consider the Rydberg states in the vicinity of static electric field $F = 3.25\:$V/cm. 
To trap the Rydberg state $\ket{r}=\ket{48D_{5/2},m_j=3/2}$ having the static dipole moment $d_r/h = 201\:$MHz/(V/cm), we couple it by a microwave field of frequency $\omega/2\pi = 42.534\:$GHz to the state $\ket{a}$ in the lower $\ket{46L>3}$ manifold and having the dipole moment $d_a/h = - 185\:$MHz/(V/cm).
The electric dipole matrix element for the transition $\ket{r} \leftrightarrow \ket{a}$ is $\wp_{ra} = 147 a_0 e$.  
The other Rydberg state $\ket{s}=\ket{49P_{3/2},m_j=3/2}$ having the static dipole moment $d_s/h = 789\:$MHz/(V/cm) is coupled by a microwave field of frequency $\omega'/2\pi = 45.543\:$GHz to the state $\ket{b}$ with the dipole moment $d_b/h = - 839\:$MHz/(V/cm) in the upper $\ket{47L>3}$ manifold.
The matrix element for the transition $\ket{s} \leftrightarrow \ket{b}$ is $\wp_{sb} = 9 a_0 e$. 
The qubit transition $\ket{r} \leftrightarrow \ket{s}$ has the frequency $\omega_{rs} - (d_r-d_a) |F|/\hbar \simeq 2\pi \times 20.488\:$GHz and a large electric dipole matrix element $\wp_{rs} = 1246 a_0 e$, which allows it to strongly couple to a resonant mode of a coplanar waveguide resonator. 

For the second example [Fig.~\ref{fig:StarkMaps}(b)], we consider higher Rydberg states in a weaker electric field of $F = 0.85\:$V/cm. 
Now, the Rydberg state $\ket{r}=\ket{59D_{5/2},m_j=3/2}$ with the dipole moment $d_r/h = 109\:$MHz/(V/cm) is coupled by a microwave field of frequency $\omega/2\pi = 22.723\:$GHz to the state $\ket{a}$ in the lower $\ket{57L>3}$ manifold with the dipole moment $d_a/h = - 113\:$MHz/(V/cm).
The electric dipole matrix element for the transition $\ket{r} \leftrightarrow \ket{a}$ is $\wp_{ra} = 200 a_0 e$.
The other Rydberg state $\ket{s}=\ket{60P_{3/2},m_j=3/2}$ with the static dipole moment $d_s = 934\:$MHz/(V/cm) is coupled by a microwave field of frequency $\omega'/2\pi = 23.174\:$GHz to the state $\ket{b}$ with the dipole moment $d_b = - 916\:$MHz/(V/cm) in the upper $\ket{58L>3}$ manifold.
The matrix element for the transition $\ket{s} \leftrightarrow \ket{b}$ is $\wp_{sb} = 9 a_0 e$.
The qubit transition $\ket{r} \leftrightarrow \ket{s}$ has smaller frequency $\omega_{rs} - (d_r-d_a) |F|/\hbar \simeq 2\pi \times 10.599\:$GHz but even larger dipole matrix element $\wp = 1945 a_0 e$, which is advantageous for strong atom-cavity coupling.

In both examples in Fig.~\ref{fig:StarkMaps}, the level separation in the corresponding hydrogenic manifold at the target fields of $F=3.25,0.85\:$V/cm is at least an order of magnitude larger than the Rabi frequencies $\Omega$ of the dressing microwave fields. 
Hence, coupling to the other nonresonant states in the hydrogenic manifold, which would potentially lead to population loss or qubit decoherence, is negligible. 

The lifetimes of all the Rydberg states in cryogenic environment of $T\lesssim 4\:$K are $1/\gamma \gtrsim 100\:\mu$s \cite{Beterov2009}. 

We finally note that the static dipole moments $d_{r,s}$ of the Rydberg states are not constant but depend on the electric field $F$ even for weak fields of a few V/cm. However, in the vicinity of the field of interest $|F| \pm 0.1\:$V/cm, $d_{r,s}$, as well as $d_{a,b}$, can be assumed constant to a very good approximation, as we did in the main text.


\begin{thebibliography}{100}
\bibitem{footnote}Due to small polarizability of the atoms in the ground electronic state, they are largely immune to mechanical forces in static, spatially inhomogeneous electric fields $F$ of moderate strengths and can be trapped by dipole forces induced by focused laser beams that form optical tweezers.
In contrast, for strongly-polarizable Rydberg states of the atoms, the mechanical forces $\bm{\nabla} (d_r|F|)$ are so large that typical optical dipole potentials cannot trap the Rydberg atoms in such inhomogeneous electric fields $F$.
\end{thebibliography}
\end{document}